\documentclass[a4paper,11pt]{article}
\usepackage{amsfonts,amssymb,url,fullpage}
\usepackage{theorem,fullpage}
\newtheorem{thesis}{Thesis}[section]
{\theorembodyfont{\rmfamily}
\newtheorem{definition}{Definition}[section]}
\begin{document}
\frenchspacing
\title{Can we debug the Universe?\thanks{An early version of this paper was read in the
``Future Trends in Hypercomputation'' Workshop held in Sheffield U.K., 11-13 September 2006.}}
\author{Apostolos Syropoulos\\
        Greek Molecular Computing Group\\
        366, 28th October Str.\\
        GR-671\ 00\ \ Xanthi, GREECE\\
        \texttt{asyropoulos@yahoo.com}}
\maketitle
\begin{abstract}
Roughly, the Church-Turing thesis is a hypothesis that describes exactly what 
can be computed by any real or feasible conceptual computing device. Generally speaking, 
the computational metaphor is the idea that everything, including the universe itself, has 
a computational nature. However, if the Church-Turing thesis is not valid, then does it make
sense to expect the construction of a computer program capable of
simulating the whole Universe? In the lights of hypercomputation,
the scientific discipline that is about computing beyond the Church-Turing 
barrier, the most natural answer to this question is: No. This note
is a justification of this answer and its deeper meaning based on
arguments from physics, the philosophy of the mind, and, of course,
(hyper)computability theory. 
\end{abstract}
\section{Introduction}
Alan Mathison Turing is considered by many as the father of modern computer science because he
set the mathematical foundation of the rough notion of computation. In the 1930s he published 
his trail-blazing paper entitled ``On Computable Numbers with an Application to the 
Entscheidungsproblem''~\cite{turing36} in which he introduced a conceptual computing 
device that now bears his name. The Turing machine is a simple and elegant device that 
consists of a scanning head that can move on top of an infinite tape, which is divided 
into cells. The scanning head can read symbols from or print symbols on each cell, but it is not
specified how it can recognize symbols, how it can erase symbols, etc. Also, the
machine's actions are predetermined by a set of rules called the controlling device. Despite the 
machine's simplicity, one can easily see that Turing machines can compute a great number of natural 
numbers and natural number functions. In fact, Turing proved that the functions computable by his machine are
precisely the {\em recursive functions} (for example, see~\cite{rogers87} for an excellent 
introduction to the theory of recursive functions).

Of course the Turing machine is not the only approach to the problem of representing/computing natural 
numbers and natural number functions. Nevertheless, a number of different approaches can be used to 
compute/describe at most the recursive functions. This observation has led Alonzo Church to formulate a 
thesis, the so-called Church-Turing thesis. This thesis, which should be better called a {\em hypothesis} 
since it is not amenable to direct proof, roughly states that only the recursive functions are computable.
And since only recursive functions are computable by a Turing machine, if one cannot construct a 
Turing machine to compute a number or a function value, then it is {\em incomputable} according to
the Church-Turing thesis.

In their long quest for a {\em mechanistic} explanation of the mind-body problem, advocates of
mechanism saw the Turing machine as a savior! Indeed, by supporting the idea that the brain is
a realization of a Turing machine, they concluded that mental states are just computational states.
Thus, one can easily ``show'' that the capabilities of anybody's mind are delimited by the
Church-Turing barrier. This approach is known in the literature as the {\em computational metaphor}.
This metaphor has been used to ``solve'' so many and different problems that is almost heretical to 
speak against its validity and its ability to really solve problems related to the functionality of the 
human brain. In addition, any idea challenging the core of the computational metaphor (i.e., any attempt 
to disconnect the capabilities of the mind from the computational capabilities of the Turing machine), is 
considered by many as an unorthodox way of thinking. 

In this context, it was extremely difficult to put forth the idea that the Church-Turing thesis is
not valid. Anyone defending such an idea would have to face the scepticism and, in many cases, the 
sarcasm of the majority of the scientific community. Nevertheless, Jack Copeland and Diane Proudfoot
made a bold move forward and openly defended {\em hypercomputation} in~\cite{copeland99}, that is,
the idea that there are both real and feasible conceptual machines that can compute more than the Turing
machine does. Not so surprisingly, Copeland and Proudfoot's hypercomputation did not harm the computational
metaphor. On the contrary, Copeland proposed in~\cite{copeland00} the so-called ``wide mechanism,'' 
that is, roughly the idea that the brain might be a hypercomputer and not just a computer. 

As was noted above, the computational metaphor was {\em applied} in so many fields, that it was just
a matter of time to be applied in physics. For example, Brian Hayes has investigated in~\cite{hayes03} the 
possibility to construct a computer program written in some ordinary computer programming language (e.g., Java
or C) capable of simulating the Universe (whatever one perceives as such). He has 
reached the conclusion that although this is a particularly difficult task, nevertheless, it is a feasible one. 
Nonetheless, this conclusion has a number of interesting consequences: Let us assume that we have at our
disposal a computer capable of running such a computer program. Then, after some time, one will
witness the emergence of intelligent life forms in this virtual universe, which, quite possibly,
one (virtual) day will construct a (virtual) computer program and a (virtual) computer capable of 
simulating their own universe, etc. I am convinced that this story is ideal for scripts for sci-fi 
movies (e.g., the movie ``The Thirteenth Floor,'' see~\url{http://www.imdb.com/title/tt0139809/} 
for movie details, is based on a very similar idea), but it has nothing to do with reality. Nevertheless, 
there are many who actually believe that we do live in a computer simulation. Going one step further, 
Nick Bostrom has suggested in~\cite{bostrom03} that the human race will be able to reach the ``posthuman'' 
stage if and only if we are living in a computer simulation! Of course, one may argue that this idea
and hypercomputation are equally absurd. However, such an argument is unfair to hypercomputation.
First, hypercomputation is about what can be computed---it is the idea that we cannot pose
artificial limits to what can be computed unless we fully understand the cosmos. In addition,
it is not some sort of obsession against the validity of the Church-Turing thesis---it is the
idea that {\em computable} does not necessarily mean {\em Turing computable}. Second, the
idea that we live in a computer simulation is closer to theology than science. After all, one
may argue that we indeed live in a ``computer simulation'' operated by God. But, this is an
entirely different problem, which should concern theologists and, then, probably, science
fiction authors. 

\paragraph{Plan of the paper} After this introduction, I will argue against the plausibility
of the idea that our universe is a computer program running on some cosmic computer (whatever 
this may mean). In particular, I will argue against the computational metaphor and
against the common belief that all physical phenomena are computational in their
nature. In the end there is a short section presenting our conclusions and ideas
for future research.

\section{Is the Human Mind a Turing Machine?}

My son knows how to add, to subtract, to multiply and to divide numbers---this is something millions, if
not billions, of people can do. Indeed, any average person can do (simple) arithmetic operations, while some 
more gifted people can perform very complicated mathematical operations without 
using paper and pencil. Nevertheless, observations like this are a proof that the human brain
(that is, the brain of an average person) has computational capabilities. But, this is certainly not a proof 
that the human brain is merely a computing device. In addition, it is interesting to see what is the
computational power of the brain, when considered as a biological machine that has computational capabilities. 
 
In the light of hypercomputation, the Church-Turing barrier does not exist and, more generally, 
there is no barrier to what can be computed simply because we do not have a full understanding of our cosmos. 
Thus, it is only possible to compare computational devices and classify their respective computational power. 
But, first, it is more than necessary to define what computation means. In general, one may define computation 
as a symbol manipulation process performed by a computational system. More specifically, a computational system 
consists of a {\em processor}, which performs the computation, and a machinery to {\em encode} data in a form manipulable by the processor. The {\em partial} or {\em final} result of the computation should be {\em decoded} 
back to a human readable form. Sometimes, the encoding and the decoding processes are trivial, but, in other 
cases, they may involve quite sophisticated procedures (e.g., think of DNA computing). Obviously, computing 
devices with similar computing power should be able to perform the same computing tasks regardless of the very 
nature of each computing task. The symbol manipulation process is a dull and unintelligent process as it
involves the processing of symbols that completely lack meaning. To make this point clear, let us consider a 
simple C program:
\begin{center}
\verb|#include <stdio.h>  |\\
\verb| #define write printf|\\
\verb|int main(){         |\\
\verb|  write("10");      |\\
\verb|}                   |
\end{center}
First of all, when this program will be executed\footnote{Strictly speaking, this program cannot be
executed. Nevertheless, one can compile it and then execute the resulting binary program or one can
directly ``executed'' by feeding it to a C interpreter.} it will print on a computer screen the string
\texttt{10}, but, obviously, as end users we do not have a way to tell whether this is a number 
or just a numeric string. In addition, by introducing a variable (i.e., \texttt{write}) to have the value
\texttt{printf}, we are making \texttt{write} a new output command. So is there any meaning
attached to the string \texttt{printf}? Not really---the compiler has been ``instruced'' to output
the various literals (numbers, strings, etc.) and/or the value of variables that are surrounded by 
parentheses and separated by commas when it encounters the following sequence of characters:
\begin{center}
\texttt{p}, \texttt{r}, \texttt{i}, \texttt{n}, \texttt{t}, and \texttt{f}.
\end{center}
In addition, what counts here is the {\em outter shape} of the letters (symbols), or, better, their ASCII or 
Unicode chararacter codes, not the sound they represent. Clearly, any a C programmer knows that he/she has
to use the string \texttt{printf} when in need to specify that something has to be printed. But this is
something no system is aware of.  Instead, every time a program is compiled, the compiler ``consults'' a table
(e.g., the header file \texttt{stdio.h} and/or the C library) and then associates this string with some machine 
code that implements the intended meaning of the string. In other words, it is clear from this discussion that 
the following definition by Stevan Harnad is in agreement with our short discussion:
\begin{definition}\label{comp:def}
Computation is an implementation-independent, systematically interpretable, symbol manipulation
process.~\cite{harnad95}
\end{definition}
From the discussion so far it is clear that meaning is more than computation. Naturally, one may object by 
saying that in science meaning is attached to objects of computation. But the physical objects, forces, etc.
are there regardless of what we compute or not. In different words, the existence of physical things is
independent from the computations that are possibly attached to them. Now, having a definition of computation, 
it is now possible to compare the computational power of a brain and the Turing machine. 

Admittedly, it is not an easy task to compare the computational power of the human mind and the Turing
machine. Nevertheless, it is known what is the computational power of the Turing machine---it
can compute at most the recursive functions. Thus, it suffices to see whether the mind is capable of
doing things no Turing machine can do. Typically, Turing incomputable problems are decision problems that 
demand an infinite number of tryouts to produce a definitive answer, while a typical Turing machine
must produce an answer in a finite number of steps. The reason why Turing machines fail to compute such 
problems is mainly because they operate in an unintelligent way. In  particular, Turing machines employ 
a brute-force search method to solve problems (i.e., the machine enumerates all possible candidates for the 
solution and checks whether each candidate satisfies the problem's statement). Clearly, this is not the 
optimal way to solve any kind of problem. Nevertheless, this is the most general approach to the problem
of checking whether an element belongs to a set. 

The statement that the equation $x^{n} + y^{n} = z^{n}$  has no non-zero integer solutions 
for all $x$, $y$, and $z$, when $n > 2$, is known in the literature as Fermat's last theorem.
As is obvious, this problem cannot be solved by a Turing machine since it has an
infinite solution space. Nevertheless, it is in principle possible to construct a Turing machine to 
verify that for some $n>0$, the equation $x^{n} + y^{n} = z^{n}$ has no non-zero solutions for all natural
numbers $x$, $y$, and $z$. Needless to say, this machine will never halt as it has to check that all $x$, $y$, 
and $z$ satisfy the given condition. Mathematically speaking, this problem is a co-recursively enumerable 
decision problem within set theory under ZF (see~\cite[p.~322]{rogers87}). In 1995, Andrew John Wiles proved 
Fermat's last theorem by using tricks from algebraic geometry. In other words, a human gave a definitive answer 
to a problem for which no Turing machine can give a definitive answer. Note that the {\em Riemann hypothesis} 
(i.e., the conjecture that the Riemann $\zeta$ function has no zeroes with real part between $0$ and $1$ other 
than those with real part equal to $\frac{1}{2}$) is similar in nature to Fermat's last theorem. This means 
that sooner or later this problem will be solved by a human. Also, note that it is possible even today to 
``construct'' a Turing machine to verify particular cases, but no Turing machine can give a definitive answer. 
These facts show that minds can hypercompute, since at least one of them can do things no Turing machine can do. 
In general, if the mind has computational capabilities beyond the Church-Turing barrier, its functional structure
must reflect these capabilities.

Peter Kugel has proposed in~\cite{kugel86} a hypercompuational model of the mind, according to which
the mind consits of four different units, one of them being a {\em trial-and-error} 
machine (see~\cite{gold,putnam}). These machines are conceptual computing devices that can solve the
Turing machine halting problem (i.e., the problem of determining whether a given Turing machine $M$ 
operating on its input $I$ will eventually stop or not) and, thus, are a feasible model of hypercomputation
as it will be demonstrated below. Typically, a trial-and-error machine is a kind of a Turing machine, 
which can be used to determine whether an element $x$ belongs to a set $X\subset\mathbb{N}$, where
$\mathbb{N}$ denotes the set of positive integers including zero. More generally, a machine
can decide whether a tuple $(x_1,\ldots,x_n)$ belongs to a relation $R\subset\mathbb{N}^{n}$. 
In the course of its operation, a machine continuously prints out a sequence of responses 
(e.g., a sequence of ones  and zeros with the intended meaning for each response). Always the last 
response is the correct one. Thus, if a machine has most recently printed ``1,'' then we know that its input
element or tuple belongs to a set or a relation, respectively, {\em unless the machine is going to change its
mind}; but we have no procedure for telling whether the machine will change again its mind or not. 
When a trial-and-error machine has printed out an {\em infinite} number of responses, after a certain point 
it will converge to a particular response and, thus, it will continuously print out the same response 
(i.e., either ``1'' or ``0''). 

Despite Kugel's reliance on hypercomputation to describe the functionality of the brain,
even though the term was not known at the time he was working on his model, the proposed model
is in spirit with Copeland's wide mechanism. Nevertheless, this model refutes Hayes's conclusion, which
is equivalent to the statement that the universe is computable.
On the other hand, this model implicitly introduces the idea that
the universe is hypercomputable. Still, this is a crude approach to the problem of the
functionality of the brain. 
ARNNs have been introduced by Hava Siegelmann and 
are a form of neural networks with hypercomputational capabilities (see~\cite{siegelmann99} for a thorough
presentation of ARNNs). And since analog computers are fundamentally more powerful than Turing machines
(see~\cite{maclennan04} for thorough discussion), Lee Albert Rubel's the-mind-as-an-analog-computer model, 
which is described  in~\cite{rubel85}, completely refutes Hayes's conclusion.

Common phenomena and everyday practices follow a trial-and-error ``behavioral pattern.'' For
example, ordinary people learn how to solve problems by unconsciously ``invoking'' a trial-and-error 
problem-solving strategy. In addition, evolution, the most important biological process is roughly a
trial-and-error process. Figuratively speaking, nature continuously creates new species by evolving
existing ones. The new species, which are more adaptable and more sophisticated compared to their predecessors,
replace their predecessors.\footnote{Recent research results, presented in~\cite{spoor07}, contest this last part 
of the commonly accepted evolutionary process by providing evidence that two particular homo species, 
a predecessor and a successor, actually co-existed longer than expected.} Nevertheless, these are not pure 
trial-and-error procedures as both are affected by
``external'' parameters (e.g., any kind of climatological change affects the evolutionary process). In this 
respect, these phenomena and practices are more general and might be characterized as {\em interactive} 
trial-and-error procedures. Such procedures are even more expressive than trial-and-error machines, since
it has been shown that interactive systems are more powerful than Turing machines (see~\cite{wegner98} for 
more details). So, one can safely conclude that evolution is an incomputable process. Furthermore, one may 
argue that any other evolutionary process is incomputable. In particular, based on the idea that culture is an 
evolutionary process, an idea that was put forth by Donald Campbell in~\cite{campbell65}, one may 
argue that culture is a process that is neither computable nor incomputable. I call any process that
is not hypercomputable a {\em paracomputational} process.

In general, I call paracomputational any process that cannot be described algorithmically. In addition, paracomputational processes cannot be simulated by any computing device, though in principle it should
be possible to simulate parts of such a process. Obviously, even a hypercomputer cannot simulate a
paracomputational process. The most striking example of a paracomputational process is an orgasm,
something that was pointed out to the author by Jaak Panksepp in a private communication. Now, I classify 
culture as a paracomputational process since it involves the communication of human minds. As was suggested by 
the famous {\em Chinese Room Argument} (CRA for short), which was put forth by John Searle in~\cite{searle80}
(but see also~\cite{wakefield03} for a fresh look to the CRA), 
the mind is not a realization of a Turing machine. As might be expected, not everybody shares this view of things 
(for example, see~\cite{cra02} for a recent collection of papers discussing the CRA). Roughly, the argument goes 
as follows: a person who neither speaks Chinese nor understands the Chinese writing system is inside a room and 
is equipped with a rule-book. He receives questions in Chinese printed on piece of paper by a person fluent in
Chinese. Using the rule-book, the person inside the room can type in responses on piece of paper.
The outsider establishes a communication with the person inside the room and thinks that he 
actually speaks Chinese, but, as we have noted, he does not. The core of this argument is that
dull symbol manipulation does not require intelligence to be carried out. As it was noted above computation
is just this---dull manipulation of meaningless symbols. Obviously, an average mind can achieve more than any 
computing device, since, for example, it is capable of understanding the meaning associated to printed words.
In other words, reading is not merely a symbol manipulation process, but something more.
So culture is a paracomputational process, since it can be viewed as an entity consisting of
minds that interact and, consequently, continuously change it.  

I have showed that even if we accept wide mechanism the Universe is incomputable. However,
the Universe is not just incomputable it is paracomputational since minds, which are parts of
this Universe, have such properties. In conclusion:
\begin{thesis}
The Universe when viewed as an entity is incomputable and paracomputable.
\end{thesis}

\section{Is the Universe a Turing Machine?}
Pierre-Simon Laplace was probably the first thinker who advocated that the Universe is computable.
Laplace's daemon is a na\"{\i}ve explanation of why it is possible to know everything that goes
on in the Universe. Essentially, Laplace's daemon is no different from Haye's computer program.
A more sophisticated theory of a computable Universe was presented by Conrad Zuse in his
monograph ``Rechnender Raum'' (Calculating Space~\cite{zuse69}), which was published in 1969. 
Zuse advocated that one should be able to answer physical questions using the tools of automata theory. 
In a nutshell, Zuse supported the idea that everything physical is granular in nature as what we perceive as 
continuoum can be divided into discrete cellular automata.\footnote{At 
\url{http://mathworld.wolfram.com/CellularAutomaton.html} one reads that ``[a] cellular automaton is a 
collection of `colored' cells on a grid of specified shape that evolves through a number of discrete 
time steps according to a set of rules based on the states of neighboring cells.''} Cellular automata are
computing devices that do not violate the Church-Turing thesis, thus, it is in principle possible to
compute all physical phenomena and, consequently, to compute the future. Naturally, this ``conclusion'' 
does not take into consideration conclusions such as these that was presented in the previous section. Of 
course, thinkers who try to show that the Universe is computable, take it for granted that everything in 
this Universe is actually a computable entity. Although, this view is definitely wrong, the general argument 
about the computational nature of the Universe is based on the assumption that space and time are granular. 

Stephen Wolfram's is another thinker who advocates that the Universe is computable. He has introduced 
in~\cite{wolfram02} the so-called {\em principle of computational equivalence}. This principle states that 
``almost all processes that are not obviously simple can be viewed as computations of equivalent 
sophistication.'' As an application of this principle, Wolfram has used it in order to explain the phenomenon 
of free will. In particular, he argues that ``[f]or even though all the components of our brains presumably 
follow definite laws, I strongly suspect that their  overall behavior corresponds to an irreducible 
computation whose outcome can never in effect be found by reasonable laws.'' From this, Wolfram goes one step
further and concludes that free will is just an illusion! Clearly, Wolfram provides no proof for what he claims.
Based on a vague principle and by assuming something about our brains, for which there is no scientific evidence,
he actually dares to conclude that free will is an illusion! Fortunately, not all physicists deal with 
such delicate issues in such a reckless way. For example, Kip Thorne in his lucid account of the nature 
of black holes points out that ``free will is a terribly difficult thing for physicists to deal with. 
We usually try to avoid it. It just confuses issues that otherwise might be lucid''~\cite[p. 507]{thorne94}. 

If cellular automata can explain free will, they can also explain life as a physical phenomenon. Indeed, Wolfram 
asserts that biological systems are the outcome of some simple {\em programs} (i.e., cellular automata). 
It is true that certain biological phenomena have an algorithmic nature. For example, the pigmentation of see 
shells is a process that can be simulated by a computer (see~\cite{meinhardt03} for more details). One the
other hand, as was shown in the previous section, evolution, which is the most important aspect of life,
is an incomputable process. Therefore, biological systems cannot possibly be the outcome of some simple
(or even quite complex!) programs. Of course Wolfram does not stop here. He actually proposes that
everything in this Universe is the outcome of a simple program! But this cannot be possibly true since
there are incomputable phenomena. On the other hand, this seems to me like a {\em digital} theology, where 
God is replaced by a computer. 

Norman Morgolus is another thinker who believes that the Universe and everything contained in it are
computers. In particular, he advocates that ``Nature is more like a spatially distributed Cellular Automaton 
than like a conventional von Neumman machine''~\cite{margolus03}. In different words, he totally agrees
with Wolfram. Clearly, if the Universe is not granular, then these theories will become groundless.
So what is known about the nature of space and time?
 
Richard Lieu and Lloyd Hillman presented in~\cite{lieu03} evidence that space
and time are not discrete. In particular, they used images of a distant galaxy
taken by the Hubble Space Telescope to directly test whether time continues to have its
usual meaning on scales of $\le t_{p}=\sqrt{\hbar G/c^{5}}\approx 10^{-44}s$, where
$t_{p}$ is the Plank time formed by the speed of light $c$, the quantum scale $\hbar$, 
and the gravitational constant $G$. Also, Roberto Ragazzoni, Massimo Turatto, and Wolfgang Gaessler
using similar techniques reached identical results that are described in~\cite{ragazzoni03}.
Nevertheless, these findings did not remain immune to challenges. In particular, Jack Ng and his collegues 
cannot conclude in~\cite{ng03} ``that modern theories of quantum gravity have been observationally ruled 
out.'' Interestingly enough, Ng and his collegues have used the techniques developed by Lieu and Hillman
to rule out two of the three theories  of quantum gravity!

If there are (strong?) indications that space and time are not granular but continuous, does 
it make any sense to try to find the {\em ultimate limits} of computation? Of course not! 
In spite of this, there are attempts to specify first the limits of computation and then to
describe the ultimate computing device. The most striking examples of such devices are
Seth Lloyd's {\em ultimate laptop}, which is described in~\cite{lloyd00}, and Ng's 
black-hole-as-a-computer, which is described in~\cite{ng01}. In particular, Ng without 
providing a definition of what is computation argues that if $\nu$ is the number of operations per unit time 
and $I$ the number of bits of information in the memory space of a black hole, then it holds that
$I\nu^{2}\lesssim t^{-2}_{p}$. The number $\bar{\nu}$ of operations per unit of time is given by
$\bar{\nu}=I\nu$. Of course, when Ng says that a black hole computes, this implies that he is in a position 
to interpret certain sequences of events that take place inside a black hole as computation. More
generally, it seems that Ng accepts {\em universal realizability} (i.e., the claim that anything can be
described as implementing a computer program). Also, it is obvious that Ng and Lloyd do accept the validity 
of the Church-Turing thesis. Now, in the light of hypercomputation, one could argue that black holes are
hypercomputers. But if black holes hypercompute, then no one can say anything about the capacity of
their ``memory'' or their ``computational'' capabilities. So, no one can say anything about ultimate
computing devices, even if universal realizability is correct. 

When one claims that a black hole computes, then of course it makes sense to claim that a chair computes,
a house computes and even the paper you are reading right now computes.  But if they compute, what do they
compute? Following, Wolfram one would argue that all things physical compute the future. Also, one
may argue that the Universe is a gigantic analog computer consisting of zillions computers, each of them
computing its own future. However, the problem with such a thesis is that it contradicts with the definition of
computation presented in Definition~\ref{comp:def}. It does make sense to ask: what are the symbols
being processed and what are the rules that are followed? In addition, one should note that even 
modern digital computers do compute just because we have designed them to do computations. And of
course when we switch off any computing device it stops processing information. So it is pure
mysticism to claim that a material object (e.g., a book, a chair, or the Universe itself) computes
something. 

For the moment let us ignore all objections that presented so far and let us accept that space and
time are granular and that the Universe is computable. Clearly, this would mean that all phenomena
that take place in the Universe will be computable. But, if one could show that there are 
incomputable phenomena taking place in the Universe, then we would have a direct proof that
the Universe is not computable---an otherwise computable Universe cannot possibly have incomputable
constituents. So are there any incomputable physical phenomena?

Marian Pour-El and Jonathan Richards have shown in~\cite{pour-el89} that non-computability
emerges naturally in wave propagation. In particular, they considered the wave propagation
\begin{displaymath}
\nabla^{2}\psi=\frac{1}{v^2}\frac{\partial^2 \psi}{\partial t}
\end{displaymath}
on compact domains and managed to find an incomputable function there. In addition, N.C.A. da Costa
and F.A. Doria have managed to express the halting function in the language of calculus (for example
see~\cite{dacosta06}). The essence of this discovery is that the halting function is in principle computable, 
if we agree that everything that can be given an explicit mathematical expression can be computed in 
some way. Moreover, Alisa Bokulich showed in~\cite{bokulich03} how to perform an infinite number of quantum 
mechanical measurements in a finite amount time, thus, implicitly showing that non-computability is an essential
ingredient of a quantum mechanical world. 

Although, I cannot accept the idea that the Universe is a finite entity, still it is pedagogically
useful to think that what the majority of the scientific community considers as Universe is a gigantic 
analog machine. But the Universe is incomputable and events take place non-deterministically:
\begin{thesis}
When viewed as an entity the Universe is an incomputable analog machine that operates 
in a non-deterministic way. 
\end{thesis}

\section{Conclusions}
I have argued why we cannot construct a computer program capable of simulating the
universe. First of all, I have explained that conscious beings have both hypercomputational and
paracomputational capabilities. Thus, it is not possible to ``(hyper-)compute'' the mind of 
conscious beings. The Universe cannot possibly be granular, that is, space and time cannot
be possibly discrete and so the Universe is analog in its very nature. Also, there are
purely physical phenomena that are incomputable, which clearly implies that the Universe
cannot be computed. In other words, we cannot debug the Universe! 
\section*{Acknowledgments}
I thank Bruno Scarpellini, Joseph Vidal-Rosset, and Francisco Antonio Doria for their many
comments and suggestions that I have used to substantially improve the arguments presented in
this paper.
\end{document}